\documentclass[doublecol]{epl2}
\title{Environmental feedback drives cooperation in spatial social dilemmas}
\shorttitle{Environmental feedback drives cooperation}

\author{Attila Szolnoki\inst{1} and Xiaojie Chen\inst{2}}
\shortauthor{Szolnoki and Chen}
\institute{\inst{1}Institute of Technical Physics and Materials Science, Centre for Energy Research, Hungarian Academy of Sciences, P.O. Box 49, H-1525 Budapest, Hungary\\
\inst{2}School of Mathematical Sciences, University of Electronic Science and Technology of China, Chengdu 611731, China}


\abstract{Exploiting others is beneficial individually but it could also be detrimental globally. The reverse is also true: a higher cooperation level may change the environment in a way that is beneficial for all competitors. To explore the possible consequence of this feedback we consider a coevolutionary model where the local cooperation level determines the payoff values of the applied prisoner's dilemma game. We observe that the coevolutionary rule provides a significantly higher cooperation level comparing to the traditional setup independently of the topology of the applied interaction graph. Interestingly, this cooperation supporting mechanism offers lonely defectors a high surviving chance for a long period hence the relaxation to the final cooperating state happens logarithmically slow. As a consequence, the extension of the traditional evolutionary game by considering interactions with the environment provides a good opportunity for cooperators, but their reward may arrive with some delay.}

\begin{document}

\maketitle

Protecting environment is often considered as a social dilemma where mutual defection can easily lead to the tragedy of the common state \cite{nordhaus_pnas10, hardin_g_s68}. In the reversed case cooperating players do not just avoid the undesired scenario but they can produce a globally improved environment that is profitable for all members. This effect is not limited to human societies, but can also be observed in microbiological systems. For example when bacteria secrete compounds for nutrient scavenging then its consequence is useful for all competitors \cite{frick_n03, hummert_jtb10, weigert_prsb17}.

We should not forget, however, that not only cooperators benefit from the improved state of the environment, but it can also elevate the temptation to defect. In other words, in a rich environment a defector gains more than a similar defector who wants to exploit others in a world where the average income is low. Accordingly, it is a more realistic approach to leave the traditional concept of fixed payoff elements and assume that these values may change in time and space. This concept assumes that the general state of environment determines the level of interactions of players, which are considered via the actual values of payoff elements \cite{weitz_pnas16}.

In this work we apply this concept by using a coevolutionary protocol where there is a feedback between the local cooperation level and payoff values of social dilemma. In particular, we introduce an adjustable coupling concerning how the actual state of environment influences the interactions of competitors and explore its consequence on the behavior of spatial systems. As we will show in what follows, the applied coevolutionary protocol has a largely unequal consequence on the evolution of strategies and cooperators will be supported by increasing the coupling with the environment. On the other hand, defectors may also benefit from improved environment, at least temporarily. Albeit they cannot escape their fate, their fight results in an unexpectedly slow relaxation that cannot be observed in the traditional model.

The remainder of this letter is organized as follows. First,
we describe the coevolutionary model that is followed by a brief discussion of the well-mixed model. Next we present our main observations for spatially structured populations, whereas lastly we summarize and discuss their
implications.

In the traditional prisoner's dilemma game we consider cooperation and defection as the two competing strategies.
The payoff elements which characterize their relations are fixed and can be interpreted in the following way: two cooperators collect $R$ reward each for mutual cooperation, a defector realizes $T$ temptation value against a cooperator, while $P$ determines the punishment for mutual defection, and finally $S$ is the income of cooperator when playing with a defector. The rank of these payoff values, namely $S<P<R<T$, ensures that it is better to defect independently of the partner's choice. Mutual defection, however, can be avoided in spatial systems where players have limited and at least temporarily fixed connections \cite{nowak_n92b}. In the last two decades several pioneering works highlighted that topology of interaction graph plays a decisive role on the high cooperation level of resulting stationary state \cite{santos_prl05, santos_prsb06, szabo_pr07, perc_pre08b, floria_pre09, fu_jtb10, roman_pre17}. We note that there are several other alternative ways to avoid the full defection state by considering more sophisticated strategies \cite{hauert_s02, vainstein_pre01, press_pnas12, yang_hx_pre09, szolnoki_pre12, javarone_csn15, suzuki_r_pre08, sicardi_jtb09}, introducing punishment or reward \cite{boyd_pnas03,brandt_pnas06,helbing_ploscb10, sigmund_pnas01, sasaki_bl14, szolnoki_epl10}.

These models, however, assume that the interactions between players of different strategies are uniform in the whole space and remain fixed in time \cite{sigmund_10}. Needless to say, this hypothesis could be oversimplified because the temptation value or the benefit of mutual cooperation may depend on site due to heterogeneous local environment. To catch the latter effect, we adopt the hypothesis when the actual state of local environment influences the quality of interaction between players, which may alter the decisions of competitors about strategy update \cite{weitz_pnas16}. This extension can be executed via a coevolutionary model where both the interaction of strategies and the resulting strategy distribution coevolve in time \cite{perc_bs10}. The feedback between the environment and individual state is based on the locally evaluated cooperation level that influences the payoff values.

In particular, we propose that $R$ and $T$ values are not fixed, but may change in time and space and their values depend on the actual cooperation level of the local community. For instance, suppose that player $i$ has $G-1$ nearest neighbors hence it is a focal player of a smaller group of $G$ members, where the number of cooperators is $n_C$. If $i$ is a cooperator then it can collect $R=R_0(1+\alpha n_C/G)$ reward from a $C-C$ link. 
If $i$ is a defector then it gets $T=T_0(1+\alpha n_C/G)$ payoff from a $D-C$ connection. Here $R_0=1$ and $T_0$ are fixed values of the traditional prisoner's dilemma game \cite{nowak_n92b} and $\alpha \ge 0$ determines the strength of feedback from environment. If $\alpha=0$ then the payoff elements become independent from the state of environment while for high $\alpha$ value there is a strong feedback between the local cooperation level and the increment of payoff elements. For simplicity we assume that the rest of payoff elements, $P$ and $S$, are fixed.

The dynamics of the strategy update which governs the microscopic evolution is based on the imitation of a more successful neighbor \cite{szabo_pr07}. During an elementary step we choose a player $i$ randomly who acquires its payoff $\Pi_i$ by playing the game with all its neighbors. Next, a randomly chosen partner of $i$, denoted by $j$, also acquires its payoff $\Pi_j$ by playing the game with all its neighbors. Player $i$ then attempts to imitate the strategy of player $j$ with the probability $w=\{1+\exp((\Pi_i-\Pi_j)/K)\}^{-1}$, 
where $K$ determines the level of uncertainty via strategy adoptions. To make our results comparable to previous findings we use $K=0.1$ noise value, but we stress that qualitatively similar results can be found for other $K$ values.

During the Monte Carlo simulations we have used at least $N=10^5$ players, but the system size was increased until $N=10^6$ players when the fraction of $C$ or $D$ players was too low. In particular, we paid special attention to avoid finite-size effect originated from the usage of small system size. Instead, we checked our results using different system sizes and accepted them only if they remained unchanged by increasing system size further. In agreement with the standard protocol during a full Monte Carlo step the above described elementary step is executed $N$ times, hence on average all players have a chance to update their states. After the successfully long relaxation, which takes typically $10^4-10^5$ Monte Carlo steps, we have averaged the fluctuating level of cooperation over another $10^4$ steps. The data were averaged over 10 independent runs. Additionally, to reach the desired accuracy in case of heterogeneous graphs we averaged the results over 100 independently generated networks.

It is worth stressing that this model does not only provide a more realistic model of the evolution of cooperation, but it also bridges the gap between games based on pair-interactions with descriptions where the social dilemma is described by public goods game-like multi-point interactions \cite{perc_jrsi13}. More precisely, the players' payoff is calculated solely based on payoff elements of pair interactions, but these values are depending on the collective behavior of the whole group.

\begin{figure}
\includegraphics[width=8.5cm]{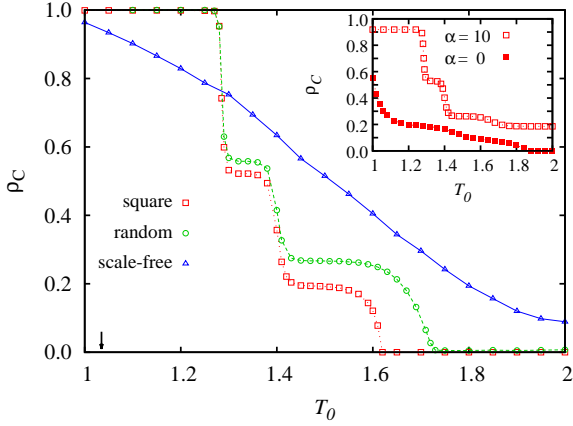}
\caption{Fractions of cooperators for three different interaction graphs at $\alpha=10$. In case of square lattice and random regular graph $z=4$ was used at $S=0$. For proper comparison in case of scale-free graph the average degree is $\langle z \rangle =4$ where $S=-0.2$ was used. An arrow at low $T_0$ value shows the typical threshold value of temptation until cooperators survive in the $\alpha=0$ case irrespectively of the graph topology. For random and scale-free topology the typical system size was $N=10^5$, while the linear system size of square lattice was $L=400$. At the vicinity of transition points we used larger system sizes as it is specified in the main text. Inset shows the results of pair-approximation in square lattice for two extreme $\alpha$ values.}
\label{A10} 
\end{figure}

Before presenting our main findings in structured populations we briefly discuss the case of unstructured populations. In a well-mixed system the fraction of cooperators can be denoted by $x$. By using this notation the average payoff of a cooperator player is $\Pi_C = x R_0 (1+\alpha x)+(1-x)S$ while the average payoff of a defector is $\Pi_D=x T_0 (1+\alpha x)+(1-x)P$. If we keep the rank $S \le P$ between the payoff elements then it is easy to see that $\Pi_D$ always exceeds $\Pi_C$. Accordingly, only the full $D$ state is evolutionary stable state. Put differently, the introduction of coevolutionary coupling between the state of environment and payoff values does not change the behavior of the traditional well-mixed model.

In spatially structured populations, however, we face to a quantitatively different situation because the heterogeneous performance of groups provides a cooperator supporting mechanism. This can already be recognized at the simplest level of multi-point approximations for square lattice topology \cite{dickman_pla86,gutowitz_pd87}.
When applying the $n$-point level of this approach, which is a dynamical version of the cluster variation method, we find a hierarchy of evolution equations for the probability distributions of configurations within a cluster of $n$ sites. (For further details and direct applications to evolutionary game systems, see \cite{szabo_jpa04, vukov_pre06, szabo_pr07}.) At $n=2$, called pair-approximation, we only have two independent variables which can determine the probability of all two-site strategy distributions. By solving numerically the equation systems we can determine the cooperation level in dependence of $T_0$ for different values of $\alpha$ parameters. In contrast to the well-mixed case here the results do depend on the strength of coupling. For two extreme $\alpha$ values the results are plotted in the inset of Fig.~\ref{A10}, which shows that coupling the present state of environment to the actual payoff values has a cooperator supporting consequence.

We can confirm this conclusion by the results of Monte Carlo simulations. In case of strong coupling ($\alpha=10$) three representative curves are plotted on the main plot of Fig.~\ref{A10} for three different interaction graphs. Here $S=0$ was applied for the square and the random regular graphs, while $S=-0.2$ was used for the scale-free graph. In the latter case $S=0$ would result in too high cooperation level even for the traditional ($\alpha=0$) case \cite{santos_prl05}, therefore we needed a negative sucker's payoff to demonstrate the difference between the traditional and coevolving models. For comparison an arrow marks the typical threshold value of $T_0$ until cooperators survive in the traditional models when spatially uniform and fixed payoff values are used.

These results suggest that there is a strong cooperator supporting consequence if we allow a feedback between the local cooperation level and the actual payoff values which characterize the interactions of strategies. While a defector gains significantly more in the neighborhood of cooperators, the positive consequence of coupling is even stronger for cooperators who can support each other more efficiently.

\begin{figure}
\includegraphics[width=4.5cm]{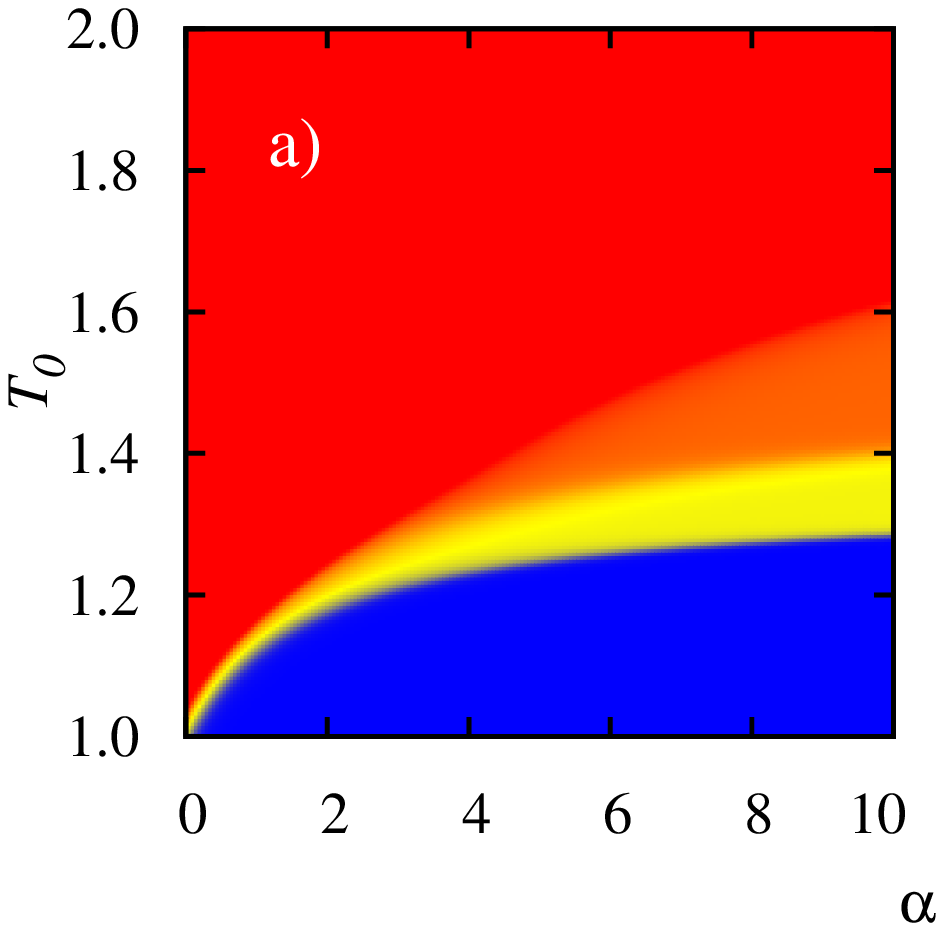}\includegraphics[width=4.5cm]{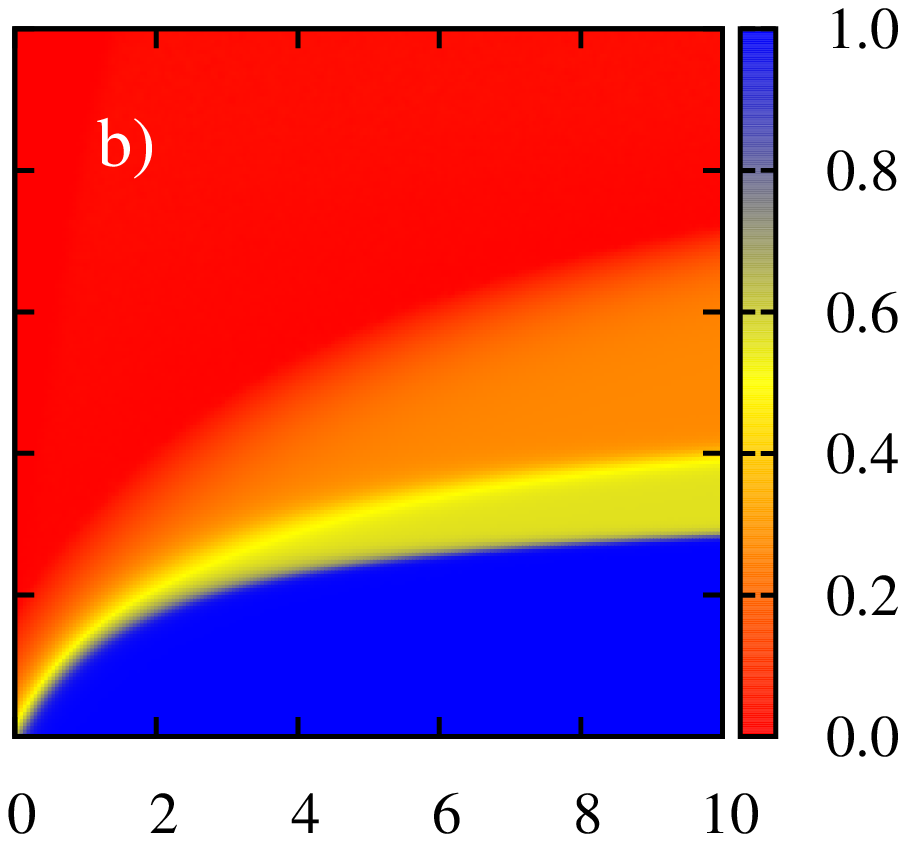}\\
\includegraphics[width=4.5cm]{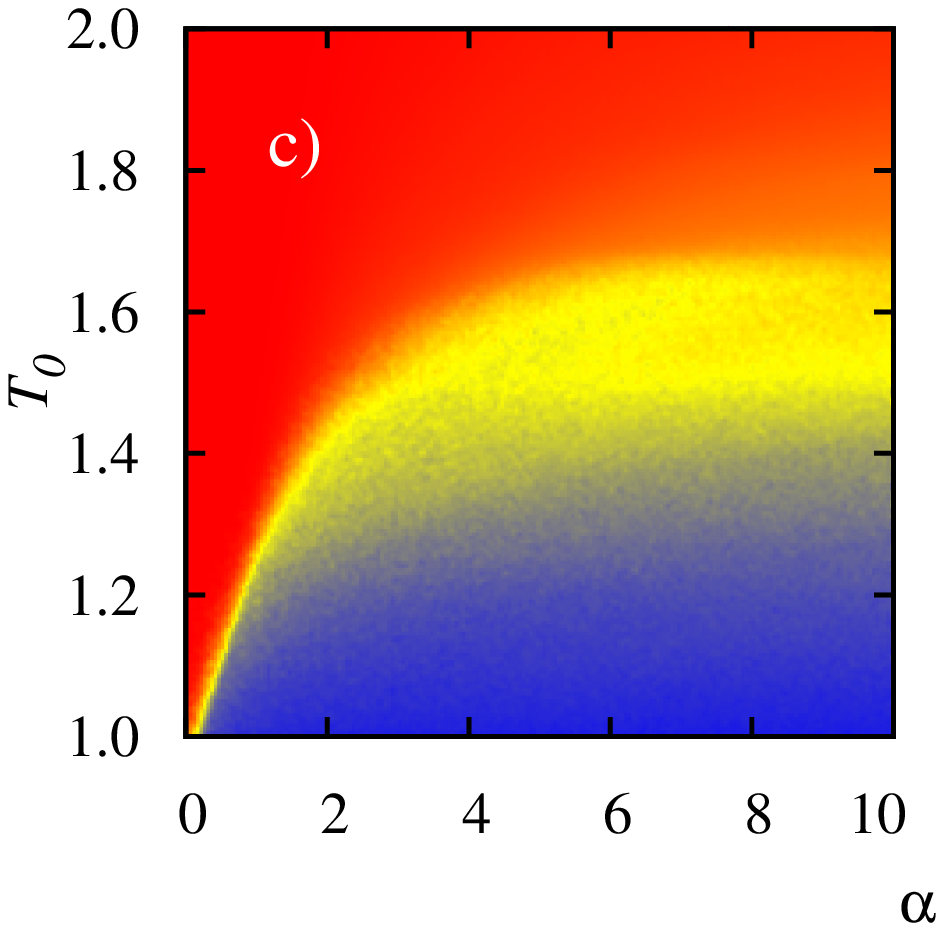}\includegraphics[width=4.5cm]{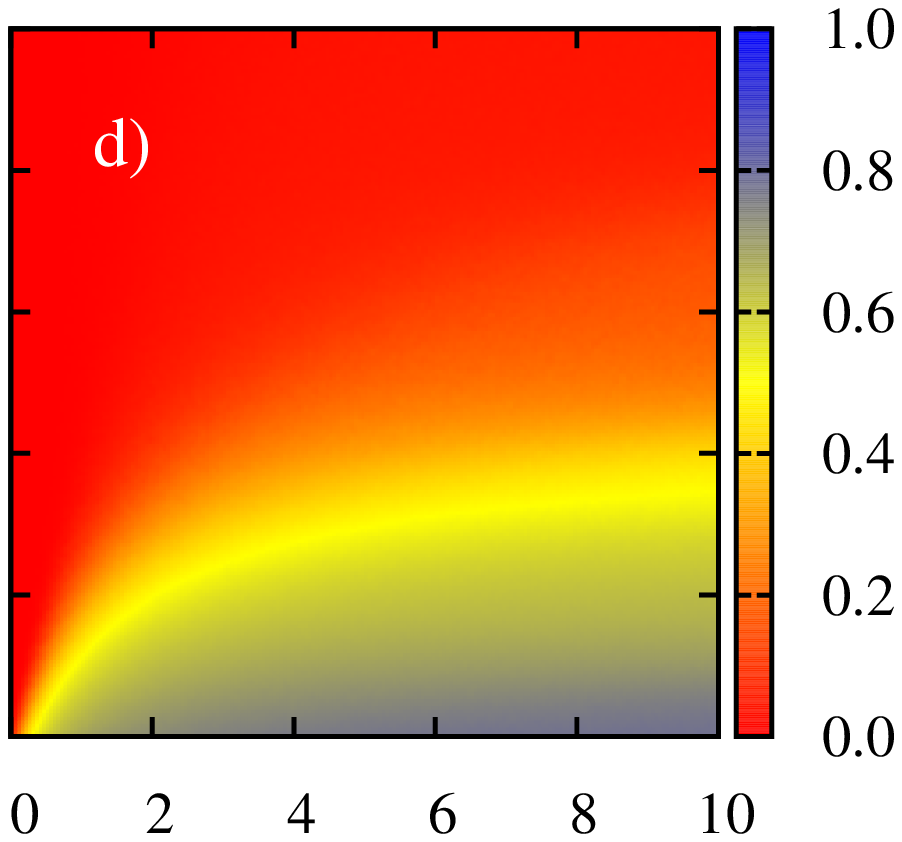}\\
\caption{Color-coded fraction of cooperators on $\alpha-T_0$ plane for four different cases. Panel~(a) shows the square lattice interaction graph at $S=0$, while panel~(b) shows the case of random regular graph using the same $S$ value. Panel~(c) shows the case of scale-free topology when $S=-0.2$ was used. Finally, panel~(d) shows scale-free topology again but here degree-normalized payoff values were applied. In the latter case $S=0$ was used again.}
\label{heat} 
\end{figure}

A more general overview about the consequence of environmental feedback can be seen in Fig.~\ref{heat} where we plotted the cooperation level in dependence of two key parameters, such as $T_0$ and $\alpha$. To demonstrate the robustness of the reported effect we have used different topologies and different ways how payoffs are calculated. Panel~(a) summarizes the results for square lattice interaction topology when $S=0$ was used. Panel~(b) shows the results when the original square lattice is fully randomized by rewiring links and introducing short cuts as it was originally described in Ref.\cite{szabo_jpa04}. In this way we introduced small-world character into the topology without changing the uniform degree-distribution of nodes. The comparison of the mentioned panels indicates that the translation invariant regularity of lattice topology does not influence the observed effect relevantly, which remains valid even if small-world like feature characterizes the host network.

\begin{figure*}
\includegraphics[width=5.5cm]{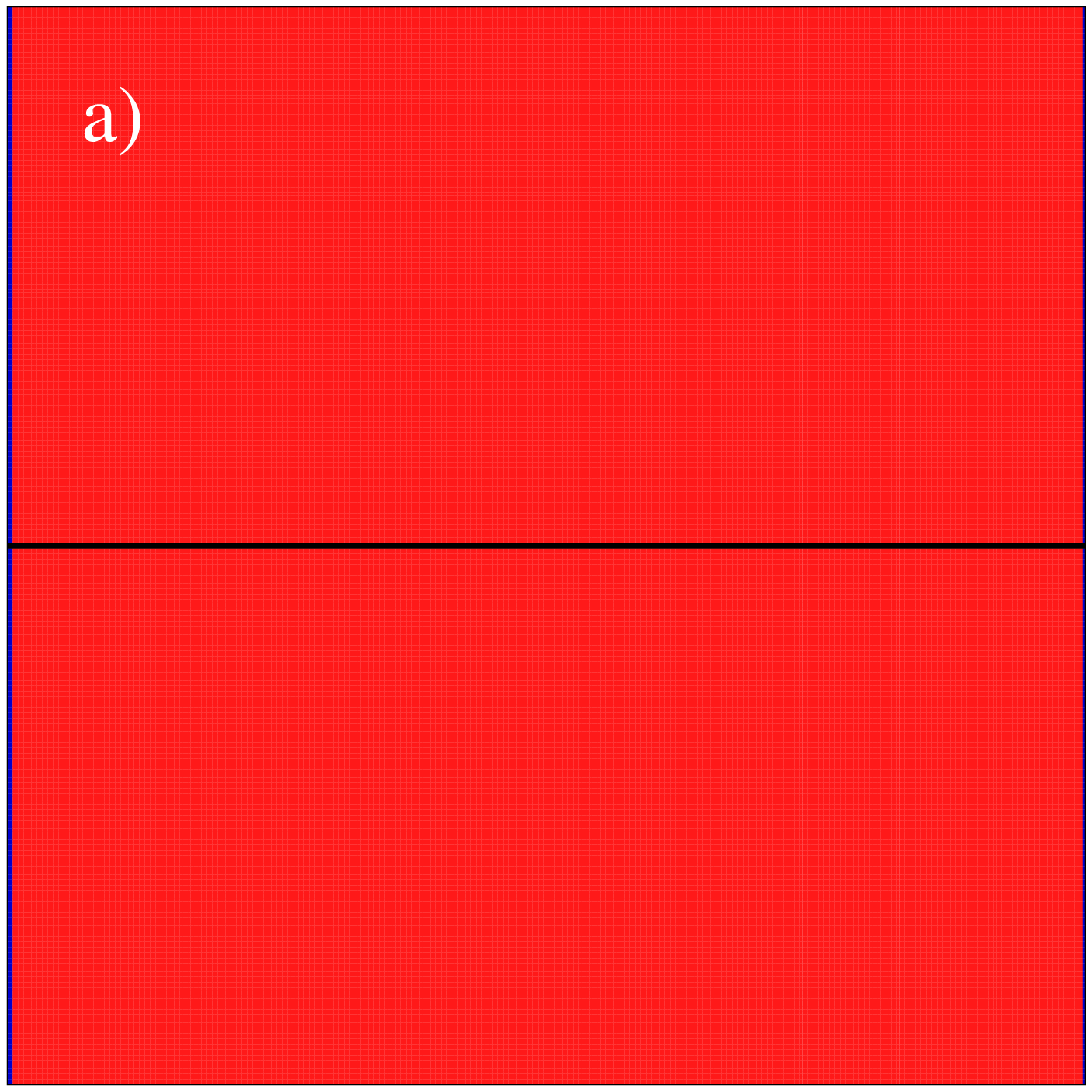}\includegraphics[width=5.5cm]{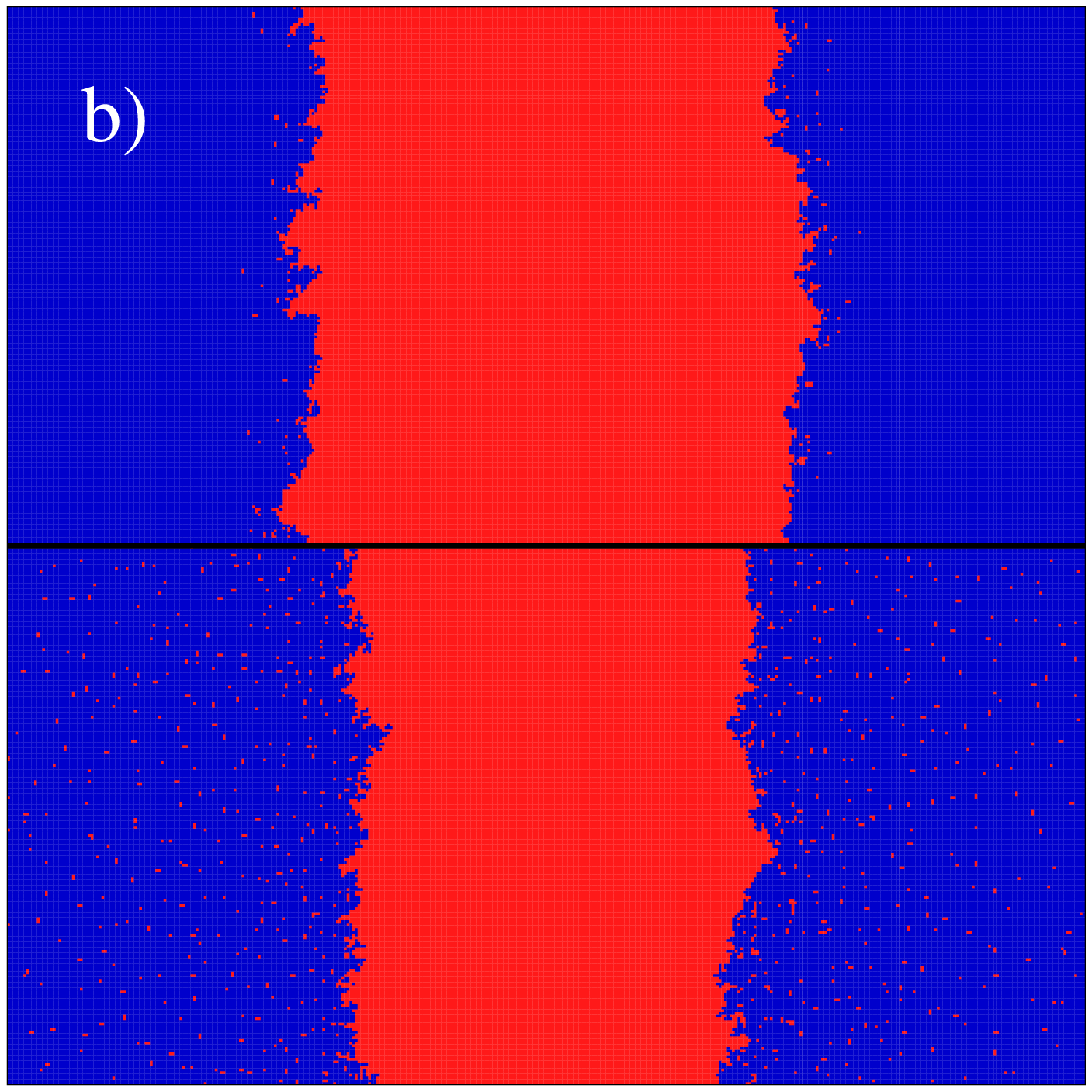}\includegraphics[width=5.5cm]{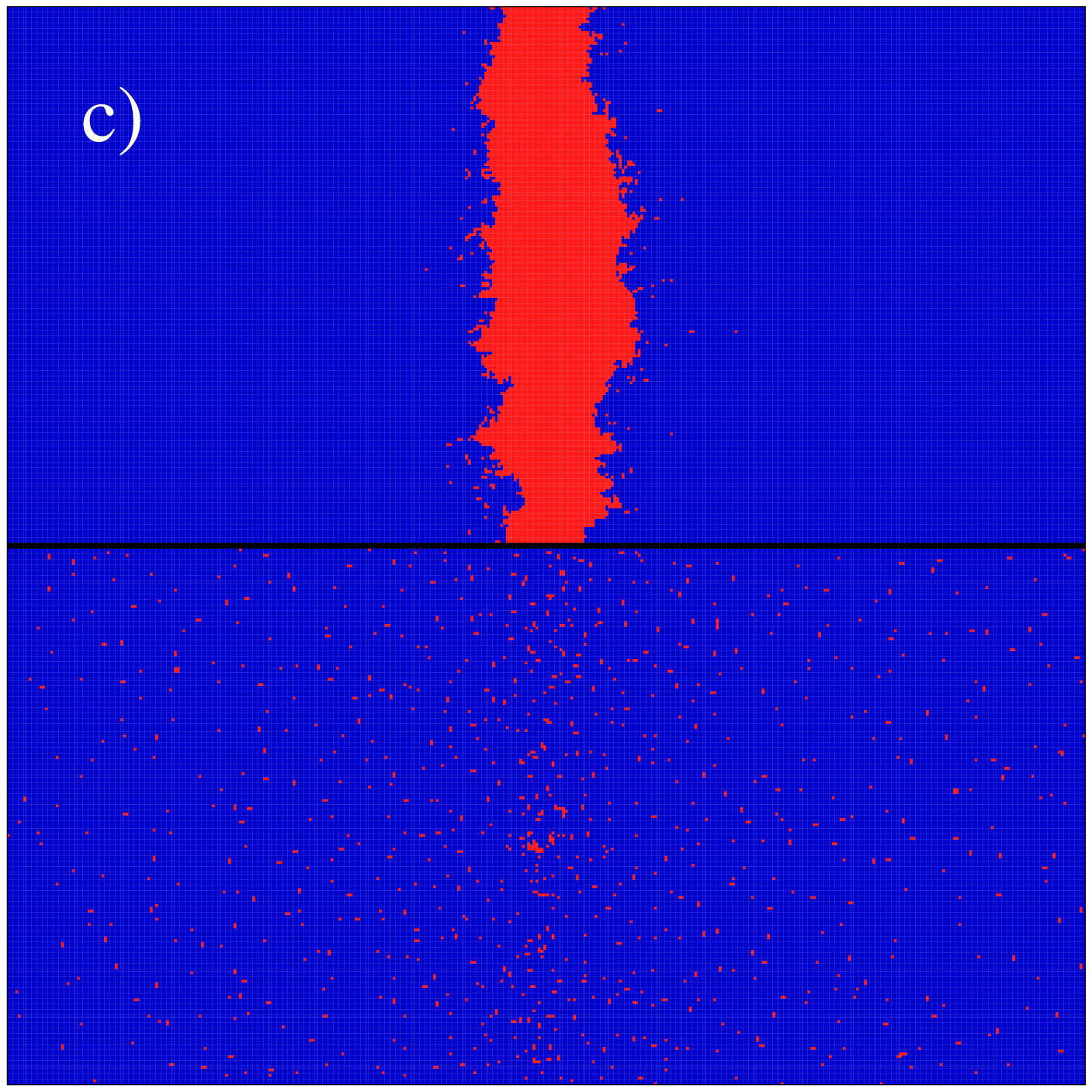}
\caption{Interface propagation stating from a prepared initial state. Here the $400 \times 400$ square lattice is horizontally divided into two parts, where in the top half we applied $\alpha =0$, while in the bottom half $\alpha=10$ is used. (To keep the distinct feedback conditions for the whole evolution, invasion across the horizontal borders is not allowed.) In both cases $T_0=0.8$ is used to ensure the final dominance of cooperators. At the beginning, shown in panel~(a), red defectors are framed by narrow strips of cooperators (the latter stripes are not visible due to large linear system size). Panel~(b) shows the states after 300 Monte Carlo steps where invasion of blue cooperators becomes visible. In the bottom half the moving interface is smoother and the invasion is faster than in the top half. After 480 MC steps, shown in panel~(c), the homogeneous defector domain already disappears in the bottom half, but the fast invasion left singular defectors behind the fronts. Finally, full cooperator state is reached for both cases but the relaxation is faster in the top half.}
\label{interface}
\end{figure*}

The positive impact of environment coupling on cooperation level remains intact if we apply largely heterogeneous degree distribution in the interaction graph. For instance, when scale-free graph is used we can observe similar effect as previously: as we increase the coupling between the environment and the applied payoff values then a higher general cooperation level can be reached and cooperators can dominate the whole system even at so high $T_0$ level which would cause their extinction in the traditional model. Note that here we had to use a significantly smaller $S=-0.2$ value because scale-free topology at $S=0$ would result in dramatically high cooperation level even for the traditional model \cite{santos_prl05, santos_pnas06}. The latter fact would make it impossible to distinguish high cooperation levels hence to illustrate the positive consequences of environmental coupling for the frequently used weak prisoner's dilemma case.
 
Turning to the last panel~(d) we still use scale-free topology but apply a conceptually different way of payoff calculation. More precisely, we normalized the payoff values of every players by their degree, which changes the cooperation level dramatically. In the latter case, when hubs cannot collect significantly higher payoff values than their neighbors or the cost of maintaining connections is considered, the general cooperation falls back to a poor level observed for graphs which are characterized by homogeneous degree distribution \cite{tomassini_ijmpc07, szolnoki_pa08, masuda_prsb07}. (This negative consequence of payoff normalization allowed us to use $S=0$ value again.) In the extended version, however, our results illustrate again that the coevolutionary coupling of environment and payoff has a positive consequence even if we use normalized payoff values.

Based on the presented observations we may conclude that the feedback mechanism, which allows coevolution of strategies and payoff values, can enhance the positive consequence of network reciprocity efficiently. This conclusion can be tested directly if we monitor how interface separating homogeneous domains evolves at different coupling rates. Figure~\ref{interface} shows such a comparison where we divided horizontally the space into two subsystems where in the top half $\alpha=0$ while in the bottom half $\alpha=10$ was used during the evolution. To maintain these differences between the subsystems we prohibited strategy invasion across the horizontal border lines (which are on the top, middle, and on the bottom of the square lattice). All the other parameter values are identical for both subsystems. We are interested in how both subsystems reach the full cooperator state therefore we used a small $T_0=0.8$ value which provides this final state even for the traditional ($\alpha=0$) case.

At the beginning we start from an almost defector state, colored by red, where their homogeneous domains are bordered only by narrow vertical stripes of cooperators. The latter are colored by blue but the width of these stripes is only 5 lattice sites which is invisible at such large linear system size ($L \times L = 400 \times 400$ was used to follow the spreading of interface properly). When evolution is launched cooperators start invading defector domains due to the small $T_0$ value. In agreement with the enhanced network reciprocity conjecture the interface moves faster and remains more regular in the bottom half where the coevolutionary protocol is applied. In the top half, where only the pure network reciprocity is at work, this domain wall is more fluctuating and the width of the defector domain shrinks slower. Interestingly, however, the faster invasion in the bottom half leaves many lonely defectors behind the front. These "snow-flake"-like pattern seem to be long life and disappear very slowly. Nevertheless, we stress that both subsystems terminate into the full cooperator state but this relaxation is qualitatively longer for the invasion which is "fastered" by coevolutionary coupling.

\begin{figure}
\includegraphics[width=8.5cm]{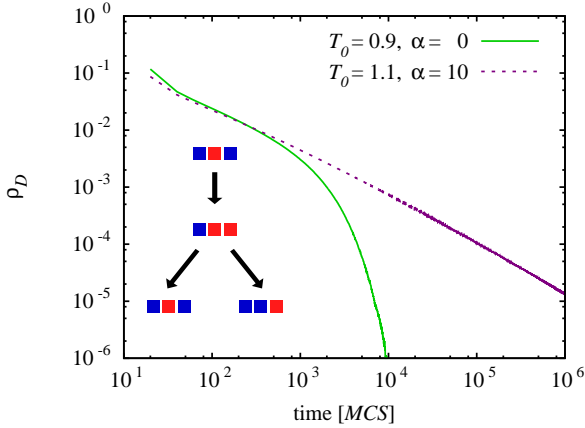}
\caption{Decay of defector concentration in two different cases where the system evolves into a full cooperator state on square lattice topology. While in the traditional model this decay is exponentially fast, in the coevolving model it happens logarithmically slow. In both cases we used $L \times L = 1000 \times 1000$ system size where we averaged over 100 independent runs. Inset explains the basic mechanism which is responsible for the slow relaxation of the coevolving model. Note that not all players are presented on the two-dimensional grid but only those whose change explains how a $D-D$ pair annihilates. Further explanation can be found in the main text.}
\label{decay}
\end{figure}

To demonstrate the qualitative differences of relaxation process to the full-$C$ state we monitored the fraction of defectors for both cases where the evolutions were launched from a random initial state. Figure~\ref{decay} shows the striking differences of how the fraction of defectors decays: 
while the relaxation is exponentially fast for the traditional model, it is logarithmically slow for the coevolutionary model (we have used log-log plots to stress the differences).

The qualitative difference of relaxation process can be easily understood based on the evolution of strategy distribution we presented in Fig.~\ref{interface}. As we already warned the reader in the introductory notes, the feedback mechanism does not only support cooperators but also defectors: a lonely defector in a highly cooperative environment can collect so high temptation value which makes it strong and provides a nice example for a neighboring cooperator player to imitate. This elementary process is illustrated in the inset of Fig.~\ref{decay}. Here we presented only those players from the grid whose strategy change explains the key steps. Importantly, when this imitation of the successful defector state is executed then the situation changes drastically. In the new circumstance, illustrated in the middle row of the inset, the old defector does not only loose a beneficial link but the emergence of a new defector will also reduce the cooperation level locally. As a result, the effective $T$ value that characterizes how much a defector can exploit from a $D-C$ link will also decay relevantly. Consequently, instead of a strong defector player we will have a weak $D-D$ pair. They become vulnerable and can be easily invaded by a neighboring cooperator player. Depending on which defector player goes extinct the position of the original $D$ remains intact or moves one step. These possible options are illustrated in the bottom row of the inset of Fig.~\ref{decay}. These elementary steps explain why a lonely defector walks randomly and why it annihilates when meeting with another defector due to the consequence of strong environment feedback.
 
The above described process can be implemented as random walking of lonely defectors who are left behind the invasion front of propagating cooperator domain. Importantly, when two lonely defectors meet randomly then they weaken each other by exactly the same reason we argued above. Technically, it means that when two random walking defectors meet then one of them is annihilated. This annihilation decreases the total number of defectors gradually which explains the unexpectedly slow relaxation we report in Fig.~\ref{decay}.

To understand the evolution of cooperation has a paramount importance in several seemingly different research disciplines \cite{pennisi_s05}. The fundamental conflict of individual and collective interests can be detected in several problems raised by psychology, sociology \cite{lomborg_asr96, sober_99}, ecology \cite{wang_xt_ec16}, biology \cite{maynard_82}, or even in cancer research \cite{ben-jacob_tm12, pacheco_if14, archetti_jtb16}. During the last decades several subtle cooperator promoting mechanisms were identified which help to understand why a single competitor gives up individual interest for a collective benefit \cite{nowak_s06, hauert_s02, szolnoki_pre15, hagel_sr16, takesue_epl17}. But most of these models ignored the fact that the collective behavior of a group might have a clear consequence on the shape of local environment, which can also influence the individual success of interacting players. We should stress that several works already concerned the difficulties originating from heterogeneous environment \cite{alonso_jsm06, chen_xj_pre09b, nishi_srep16, gracia-lazaro_csf13}, but they generally assumed a stable background where the strategy choice of competing strategies has no direct consequence on the change of environment.

A more subtle approach is when we assume that the players act influences the state of the environment that has a direct consequence on the success how players interact with each other. This idea can be captured by means of a coevolutionary model where not only strategies but also payoff elements may evolve \cite{weitz_pnas16}. By following this research path we assume that the cooperation level of a local community directly determines the temptation value and the reward of mutual cooperation. This feedback can support cooperator groups, but it can also provide a stronger temptation to choose defection. In this way we are facing with the original dilemma of the competing strategies. 

We argue that in a well-mixed population this feedback has no observable consequence on the competition of strategies and the system terminates into the well-known globally defector state. In structured populations, however, cooperators gain more and a significantly improved cooperation level can be reached for intensive coupling. In the latter case even a full cooperator state can be reached at a temptation value which would ensure a complete defection in the feedback-free traditional evolutionary game independently of the applied interaction topology. On the other hand, beside the unambiguous positive effect of environment-feedback there is an interesting consequence on the relaxation dynamics how the above mentioned cooperator state is reached. Due to enhanced temptation lonely defectors die out slow because they remain vital without followers. Their temporary success will result in an effective random walk which eventually eliminates them.

Our work highlights that considering coevolutionary systems  \cite{perc_bs10, wu_t_pcbi17, richter_bs17, stewart_pnas14} may provide not only more realistic modeling of nature but can also offer several unanticipated outcomes. These consequences are expected to work mostly in spatial systems where the spatially heterogeneous background prevents averaging, hence diminishing strategy-specific consequence of feedback mechanisms. In future studies it will be interesting to see how the concept of varying payoff values provides other new mechanisms of promoting the evolution of cooperation.

\begin{acknowledgments}
This research was supported by the Hungarian National Research Fund (Grant K-120785) and by the National Natural Science Foundation of China (Grants No. 61503062).
\end{acknowledgments}

\end{document}